# Self-consistent calculations of phonon scattering rates in GaAs transistor structure with one-dimensional electron gas

A.V. Borzdov, D.V. Pozdnyakov[*], V.O. Galenchik, V.M. Borzdov, and F.F. Komarov

*Radiophysics and Electronics Department of Belarus State University,
Nezavisimosty av.4, 220050 Minsk, Belarus*

**PACS** 72.10.Di, 73.21.Hb, 73.40.Kp, 73.63.Nm

**Abstract**
Self-consistent calculations of acoustic and polar optical phonon scattering rates in GaAs quantum wire transistor structures were carried out with account of collisional broadening. The influence of the gate bias on the scattering rates was examined, too. It was shown that in order to treat the scattering rates rigorously it is important to search for electron energy levels by means of the self-consistent solution of Schrodinger and Poisson equations and to take into account the collisional broadening.

Modern technologies in micro- and nanoelectronics allow different semiconductor device structures of very small sizes and various geometry to be created. In particular, high-speed field effect GaAs transistors with one-dimensional electron gas (1DEG) or, in other words, quantum wire transistors [1, 2] have been successfully created and actively investigated. It is supposed that on the basis of these transistors a new generation of high-speed GaAs ULSI circuits can be produced.

It is known that the efficient design of such devices today is a very hard problem without numerical simulation of their electrical characteristics. In this regard, one of the most promising simulation methods is the Monte Carlo one [3]. Its correct implementation is possible when the rigorous expressions for the scattering rates for all dominant scattering mechanisms in the simulated structure are known. In this connection it is necessary to note that acoustic and polar optical phonon scattering are two of the important scattering mechanisms in GaAs structures with 1DEG [4–7].

Several works were devoted to the calculation of acoustic and polar optical phonon scattering rates (A&POPSR) in quantum wires. Particularly, in the papers [5, 7] expressions for the A&POPSR in the rectangular quantum wires were proposed taking into account the collisional broadening (in case of intersubband scattering being the same as the broadening of the energy levels) in the first order approximation. It was shown that the functions of A&POPSR versus electron kinetic energy have not any singularity points. The latter is especially important for the direct use of the derived expressions in Monte Carlo procedures as it naturally eliminates the tendency to infinity of A&POPSR in singularity points related to the features of the density of states in quantum wires. Besides, the rigorous incorporation of the phonon scattering into the charge transport simulation by means of Monte Carlo method requires the exact knowledge of electron subband energy levels in the considered quantum well as they directly enter into the expressions for A&POPSR. These energy levels in real device structures with an arbitrary quantum well can be found only by means of numerical self-consistent solution of the corresponding Schrodinger and Poisson equations.

In this work we present the results of calculation of A&POPSR in the GaAs-transistor structure (see Fig. 1), with taking into account the collisional broadening caused by only acoustic and polar optical phonon scattering, and examine the influence of the gate bias on these rates.

In general form, the acoustic phonon scattering rate $W_{ij}^A$ from an initial subband $\{i, j\}$ to all the final subbands $\{l, k\}$ in elastic approximation (such approximation is proved for quantum wires in Ref. [8]) can be presented as [4, 5, 8, 9]

$$W_{ij}^A(E, \Gamma_{ij}) = \frac{B_{ac}^2 k_B T \sqrt{2m^*}}{\hbar^2 v^2 \rho} \sum_{l=1}^{\infty} \sum_{k=1}^{\infty} D_{ij}^{lk}(E, \Delta E_{ij}^{lk}, \Gamma_{ij}) \iint |\psi_{ij}(x,y)|^2 |\psi_{lk}(x,y)|^2 \, dx dy, \quad (1)$$

$$D_{ij}^{lk}(E, \Delta E_{ij}^{lk}, \Gamma_{ij}) = \Theta(E - \Delta E_{ij}^{lk}) \sqrt{\frac{2\Gamma_{ij} + 2\sqrt{4(E - \Delta E_{ij}^{lk})^2 + \Gamma_{ij}^2}}{4(E - \Delta E_{ij}^{lk})^2 + \Gamma_{ij}^2}}, \quad (2)$$

where $k_B$ is the Boltzmann constant; $\hbar$ is the Planck constant; $m^*$ is the electron effective mass; $\rho$ is the mass density of GaAs; $v$ is the sound velocity in GaAs; $B_{ac}$ is the acoustic deformation potential; $T$ is the temperature; $\Theta$ is the unit step function; $E$ is the electron kinetic energy; $\psi_{ij}(x,y)$ and $\psi_{lk}(x,y)$ are the wave functions of the initial state with subband energy level $E_{ij}$ and quantum numbers $\{i, j\}$ and of the final one with subband energy level $E_{lk}$ and the quantum numbers $\{l, k\}$, respectively; $\Gamma_{ij}$ is the collisional broadening factor which characterizes the electron energy uncertainty caused by all possible scattering mechanisms in the state with quantum numbers $\{i, j\}$ [9] (in the given case we are considering only the acoustic and polar optical scattering mechanisms); $\Delta E_{ij}^{lk} = E_{lk} - E_{ij}$. In Eqs. (1) and (2) $\{i, j\}$ and $\{l, k\}$ are the quantum numbers describing the initial and final electron states in the directions $X$ and $Y$, respectively.

---

[*] *pozdnyakov@bsu.by*

For the polar optical phonon scattering rates with phonon emission and forward electron scattering $[W_f^e]_{ij}^{PO}$, phonon emission and backward electron scattering $[W_b^e]_{ij}^{PO}$, phonon absorption and forward electron scattering $[W_f^a]_{ij}^{PO}$, and phonon absorption and backward electron scattering $[W_b^a]_{ij}^{PO}$ are valid the following equations [6, 7, 9]:

$$[W_{f,b}^{e/a}]_{ij}^{PO}(E, \Gamma_{ij}) = \sum_\alpha \frac{e^2 \omega_\alpha \sqrt{2 m_\alpha^*}}{4 \hbar L_x L_y} \left( \frac{1}{\varepsilon_\alpha^\infty} - \frac{1}{\varepsilon_\alpha} \right) \left( n_\alpha + \frac{1}{2} \pm \frac{1}{2} \right) \sum_{l=1}^\infty \sum_{k=1}^\infty D_{ij}^{lk}(E, \Delta E_{ij}^{lk} \pm \hbar \omega_\alpha, \Gamma_{ij}) \times$$

$$\sum_{p=1}^\infty \sum_{r=1}^\infty \frac{\left| \int_0^{L_x} \int_0^{L_y} S_\alpha(x,y) \psi_{ij}^*(x,y) \psi_{lk}(x,y) \sin(p\pi x / L_x) \sin(r\pi y / L_y) \, dx dy \right|^2}{[q_{f,b}^{e/a}]_\alpha^2 + (p\pi / L_x)^2 + (r\pi / L_y)^2}, \quad (3)$$

$$[q_f^{e/a}]_\alpha = \frac{\sqrt{2 m_\alpha^* E}}{\hbar} - \frac{\sqrt{2 m_\alpha^* (E - \Delta E_{ij}^{lk} \mp \hbar \omega_\alpha)}}{\hbar}, \quad (4)$$

$$[q_b^{e/a}]_\alpha = \frac{\sqrt{2 m_\alpha^* E}}{\hbar} + \frac{\sqrt{2 m_\alpha^* (E - \Delta E_{ij}^{lk} \mp \hbar \omega_\alpha)}}{\hbar}, \quad (5)$$

where $e$ is the magnitude of electron charge; $\omega$ is the cyclic frequency of the polar optical phonon; $\varepsilon^\infty$ and $\varepsilon$ are the optic and static dielectric permittivity of the matter, respectively; $n$ is the Bose-Einstein distribution function; $L_x$ and $L_y$ are the thickness and width of the transistor structure, respectively; $\alpha$ is the subscript describing the matter, i.e. $\alpha = \{\text{GaAs, AlAs}\}$; $S$ is the function satisfying the following conditions: $S_{\text{GaAs}}$ is equal to 1 in GaAs and equal to 0 in AlAs, $S_{\text{AlAs}}$ is equal to 0 in GaAs and equal to 1 in AlAs.

The cross-section of the structure in the plane $XY$, which is considered in this work, is schematically presented in Fig. 1.

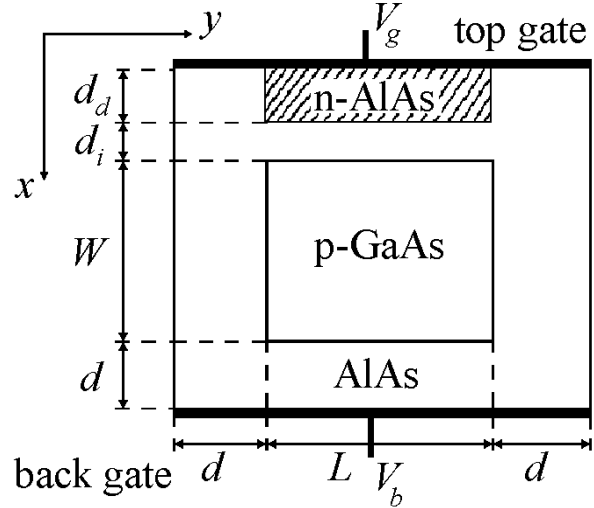

**Figure 1.** The cross-section of the GaAs transistor structure with 1DEG.

The structure represents the GaAs quantum wire with the thickness of $W = 15$ nm and width of $L = 15$ nm, surrounded with the selectively doped AlAs. The doped AlAs layer with the thickness of $d_d = 10$ nm is separated from p-GaAs by the undoped spacer layer of AlAs with thickness of $d_i = 5$ nm. We assumed that the parameter $d$ is equal to 15 nm, the donor impurity concentration in AlAs is equal to $5 \times 10^{23}$ m$^{-3}$, and the background acceptor impurity concentration in GaAs is equal to $10^{20}$ m$^{-3}$. It is supposed that the control gate bias $V_g$ is applied to the top gate relatively to the back gate.

The electron states in such a structure can be found by solving the Schrodinger equation which in this case takes the form [10]

$$-\frac{\hbar^2}{2} \nabla \left( \frac{1}{m^*(x,y)} \nabla \psi_{ij}(x,y) \right) + V(x,y) \psi_{ij}(x,y) = E_{ij} \psi_{ij}(x,y), \quad (6)$$

where the potential energy term $V$ can be written as [10, 11, 12]

$$V(x,y) = V_h(x,y) + V_{xc}(x,y) - e\varphi(x,y). \quad (7)$$

In the above equations $\varphi$ is the electrostatic potential of the self-consistent field; $V_h$ is the heterostructure potential describing the gap of conductivity zone on the GaAs/AlAs interface; $V_{xc}$ is the correlation potential. Potentials $V_h$ and $V_{xc}$ were treated similarly to [12]. The potential $\varphi$ can be found from the Poisson equation [1, 10, 11]

$$\nabla(\varepsilon(x,y)\nabla\varphi(x,y)) = en_e(x,y) - \rho_{depl}(x,y), \quad (8)$$

where $n_e$ is the electron gas concentration (see, for example, [10]) and $\rho_{depl}$ is the ionized impurity charge density.

In order to find the electron energy levels $E_{ij}$ and the wave functions $\psi_{ij}(x,y)$ in the considered device structure, the system of Eqs. (6) and (8) must be solved self-consistently [10, 11]. Then the A&POPSR determined by the formulae (1) – (5) can be calculated in accordance with [9] using the following equation:

$$\Gamma_{ij} = \hbar/\tau_{ij}^{\Sigma} = \hbar\left(W_{ij}^{A} + [W_f^e]_{ij}^{PO} + [W_b^e]_{ij}^{PO} + [W_f^a]_{ij}^{PO} + [W_b^a]_{ij}^{PO}\right). \quad (9)$$

It is necessary to solve Eq. (9) numerically for any value of $E$ in the considering energy interval.

As an example, in Fig. 2 the A&POPSR for the ground quantum state ($i = j = 1$) versus $V_g$ with taking into account three excited states are plotted. The back gate bias $V_b$ was constant during the calculation and set to 0.

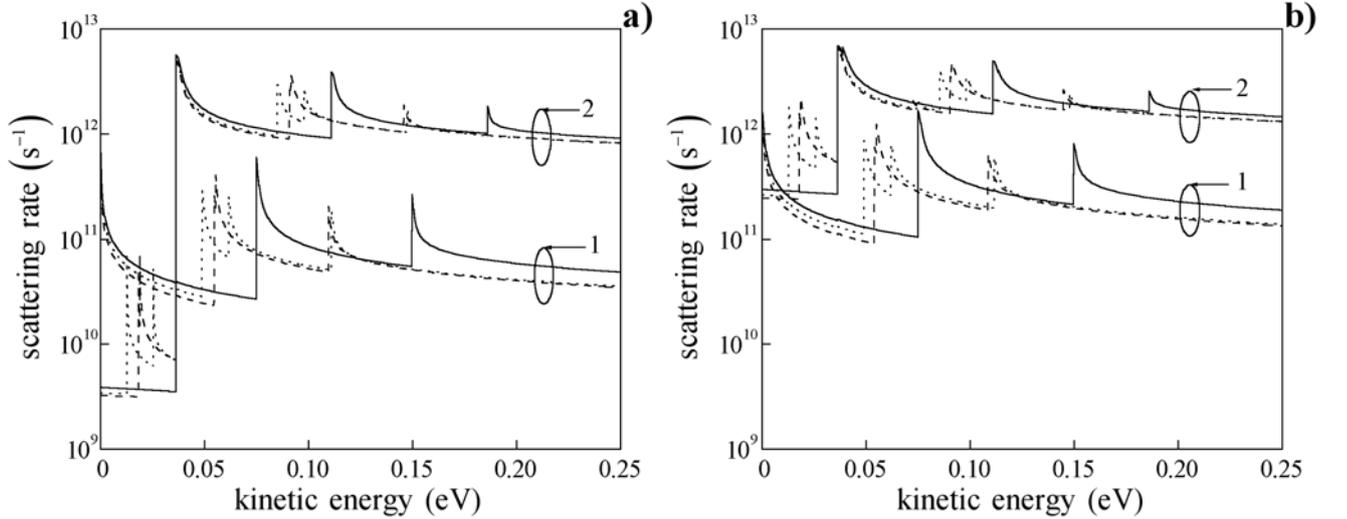

**Figure 2.** A&POPSR calculated versus electron kinetic energy for the GaAs transistor structure at $T = 77$ K (a) and $T = 300$ K (b).
*Curves* 1 – acoustic phonon scattering; *curves* 2 – polar optical phonon scattering.
*Solid lines* – the sine-like approximation of the wavefunctions for the rectangular potential well with infinitely high walls; *dashed lines* – the self-consistent results at $V_g = 0$ V; *dotted lines* – the self-consistent results at $V_g = 0.4$ V.

The figure shows that the gate bias has a great influence on the values of the A&POPSR. Moreover, the figure also shows that the use of the sine-like wavefunctions without the self-consistent procedure considerably changes the behavior of the scattering rate dependences on electron kinetic energy. Thus, it may give improper results while simulating the electron transport in the conductive channel of transistor structures. Consequently, for the exact description of the electron transport in devices of this kind, it is necessary to calculate the A&POPSR using the self-consistently calculated wavefunctions and subband energy levels.

In conclusion it should be noted that our purpose is the further study of other important scattering mechanisms and investigation of their influence on electron transport in real transistor structures.